\documentclass{XrU2005}
\include{graphicx}
\title{Supernova Remnant 1987A: High Resolution Images and Spectrum from 
Chandra Observations}
\author[1]{S. Park}
\author[2,4]{S. A. Zhekov}
\author[1]{D. N. Burrows }
\author[1]{J. L. Racusin}
\author[2]{R. McCray}
\author[3]{K. J. Borkowski}
\affil[1]{Department of Astronomy and Astrophysics, 525 Davey Lab, 
Penn State University, University Park, PA. 16802, USA}
\affil[2]{JILA, University of Colorado, Boulder, CO. 80309, USA}
\affil[3]{Department of Physics, North Carolina State University, Raleigh, NC. 27695, USA}
\affil[4]{Space Research Institute, Sofia, Bulgaria}

\begin{document}


\maketitle

\begin{abstract}
We report on the morphological and spectral evolution of SNR 1987A from 
the monitoring observations with the Chandra/ACIS. As of 2005, the 
X-ray-bright lobes are continuously brightening and expanding all around 
the ring. The softening of the overall X-ray spectrum also continues. 
The X-ray lightcurve is particularly remarkable: i.e., the recent soft X-ray 
flux increase rate is significantly deviating from the model which successfully
fits the earlier data, indicating even faster flux increase rate since early
2004 (day $\sim$6200). We also report results from high resolution spectral 
analysis with deep Chandra/LETG observations. The high resolution X-ray line 
emission features unambiguously reveal that the X-ray emission of SNR 1987A 
is originating primarily from a ``disk'' along the inner ring rather than from 
a spherical shell. We present the ionization structures, elemental abundances, 
and the shock velocities of the X-ray emitting plasma.
\end{abstract}

\section{Introduction}

The continuous development of optically bright spots around the inner ring 
of supernova (SN) 1987A indicates that the blast wave shock is approaching 
the dense circumstellar material (CSM) that was produced by the massive 
progenitor's equatorial stellar winds. In such a case, the blast wave shock 
front should decelerate by interacting with the dense CSM, and may produce 
significant soft X-ray emission. We have been monitoring the evolution of 
SN 1987A with the Chandra X-Ray Observatory since 1999, and found that the 
soft X-ray flux from SN 1987A has indeed been rapidly increasing (Burrows et al. 
2000; Park et al. 2002;2004). The significant interaction of the blast wave
with the dense CSM therefore signals the birth of a supernova remnant (SNR) 1987A.
We successfully described the composite soft X-ray (0.5$-$2 keV) light curve, 
obtained with the ROSAT and the Chandra between 1990 and 2003, using a simple 
model that assumes a constant-velocity shock interacting with an exponential 
ambient density profile (Park et al. 2004;2005a). We report here the latest 
development of the X-ray light curves of SNR 1987A, in which we find a 
significant deviation of the soft X-ray flux from the previous model predictions 
since day $\sim$6200. 

In addition to the regular monitoring observations, we have performed two deep 
Chandra gratings observations of SNR 1987A. The early observations with the High 
Energy Transmission Gratings Spectrometer (HETG) indicated a high-velocity shock 
(v $\sim$ 3400 km s$^{-1}$) in which the X-ray emitting plasma is in electron-ion
non-equilibrium (Michael et al. 2002). SNR 1987A was, however, faint and the
detected photon statistics were limited. Recently, we performed follow-up 
deep observations of SNR 1987A with Low Energy Transmission Gratings Spectrometer
(LETG) (Zhekov et al. 2005a). With the SNR being $\sim$8 times brighter and the 
exposure $\sim$3 times deeper than the early HETG observations, the new LETG 
data provide good photon statistics and allow us, for the first time, to measure 
the individual line profiles and line ratios from SNR 1987A. We present here 
some first results from these LETG data.

\section{Observations}

As of 2005 July, we have performed a total of eleven monitoring observations
of SNR 1987A with Advanced CCD Imaging Spectrometer (ACIS) aboard Chandra
(Table~1). We also have performed two deep gratings observations in 1999
October and 2004 August-September (Table~1). The data reduction of these
observations have been described in literatures (Burrows et al. 2000;
Park et al. 2005c; Michael et al. 2002; Zhekov et al. 2005a).

\begin{table*}
  \begin{center}
    \caption{Chandra Observations of SNR 1987A.}\vspace{1em}
    \renewcommand{\arraystretch}{1.2}
    \begin{tabular}[h]{ccccc}
      \hline
      Observation ID & Date (Age$^1$)  & Instrument & Exposure (ks) & Counts \\
      \hline
      124+1387$^2$ & 1999-10-6 (4609) & ACIS-S + HETG & 116.1 & 690$^3$ \\
      122 & 2000-1-17 (4711) & ACIS-S3 & 8.6 & 607 \\
      1967 & 2000-12-07 (5038) & ACIS-S3 & 98.8 & 9030 \\
      1044 & 2001-4-25 (5176) & ACIS-S3 & 17.8 & 1800 \\
      2831 & 2001-12-12 (5407) & ACIS-S3 & 49.4 & 6226 \\
      2832 & 2002-5-15 (5561) & ACIS-S3 & 44.3 & 6427 \\
      3829 & 2002-12-31 (5791) & ACIS-S3 & 49.0 & 9277 \\
      3830 & 2003-7-8 (5980) & ACIS-S3 & 45.3 & 9668 \\
      4614 & 2004-1-2 (6157) & ACIS-S3 & 46.5 & 11856 \\
      4615 & 2004-7-22 (6359) & ACIS-S3 & 48.8 & 17979 \\
      4640+4641+5362 & 2004-8-26$\sim$9-5 & ACIS-S + LETG & 289.0 & 16557$^3$ \\
      +5363+6099$^2$ & ($\sim$6400) & & & \\
      5579+6178$^2$ & 2005-1-12 (6533) & ACIS-S3 & 48.3 & 24939 \\
      5580+6345$^2$ & 2005-7-14 (6716) & ACIS-S3 & 44.1 & 27048 \\
      \hline \\
      \end{tabular}
    \label{tab:table}
\end{center}
\vspace{-5mm}
$^1$ Day since SN.\\
$^2$ These observations were splitted by multiple sequences which were combined
for the analysis.\\
$^3$ Photon statistics are from the zeroth-order data.\\ 
\end{table*}


%


%

\section{X-Ray Images}

\begin{figure*}
\centering
\includegraphics[width=0.9\linewidth]{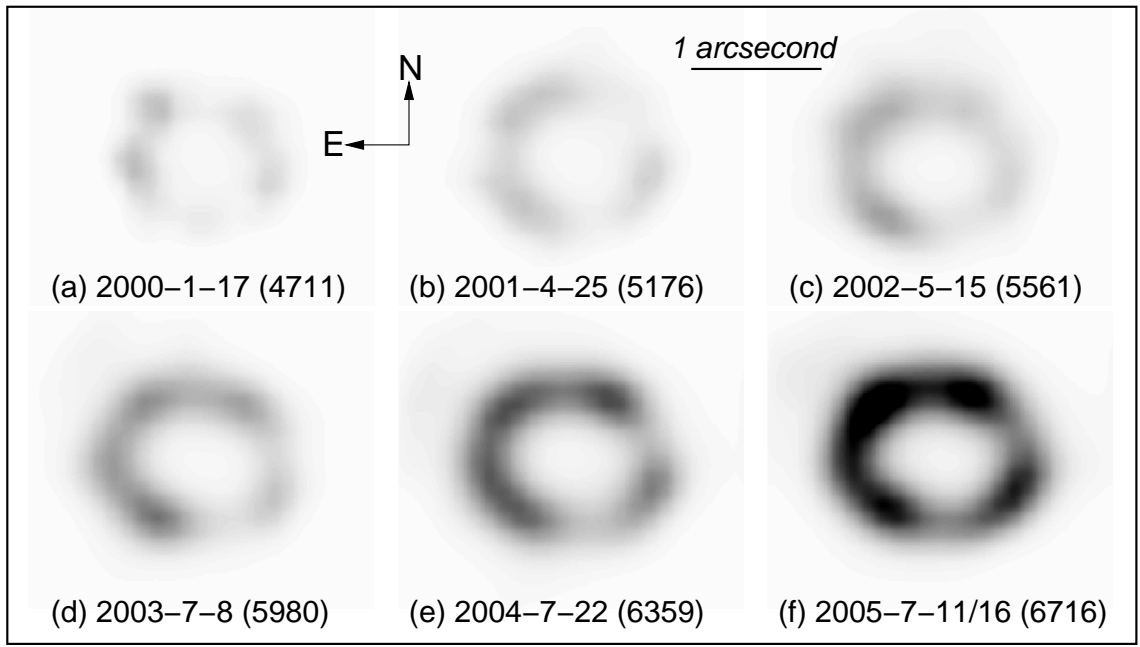}
\caption{The 0.3$-$8 keV band ACIS images of SNR 1987A. These images have been
deconvolved with the detector PSF and then smoothed for the purpose of display
following methods described in literatures (Burrows et al. 2000; Park et al.
2002). Darker gray-scales are higher intensities.
\label{fig:im}}
\end{figure*}

X-ray images of SNR 1987A from the ACIS observations are presented in Figure~1. 
Figure~1 shows images from six epochs which are separated roughly by $\sim$1 
year from each other. The continuous brightening of the X-ray emission from the 
SNR is evident. The significant brightening of the remnant makes the
X-ray morphology now a complete ring. This X-ray morphology suggests
that the blast wave shock is now engulfing the entire inner ring rather
than small clumps of the dense CSM.

%
%

\section{X-Ray Flux}

\begin{figure}
\centering
\includegraphics[angle=-0,width=\linewidth]{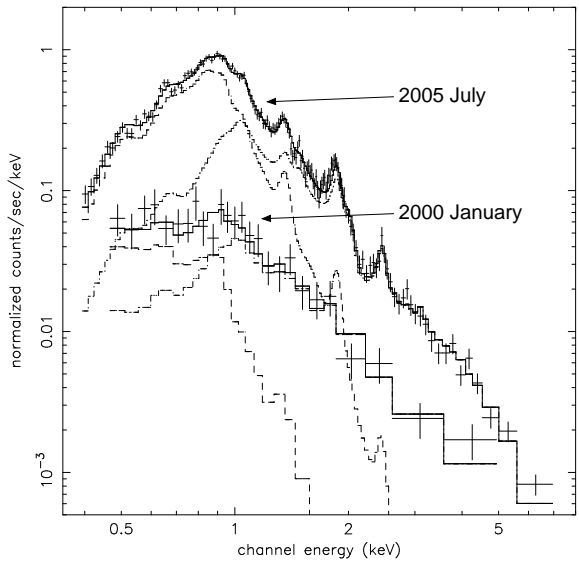}
\caption{X-ray spectrum of SNR 1987A as taken with the ACIS-S3 in 2000 
January and 2005 July. The best-fit two-shock models are overlaid on each
spectrum.
\label{fig:ccd}}
\end{figure}

Figure~2 shows the ACIS spectrum of SNR 1987A from two epochs of 2000 January
and 2005 July. The overall increase of the X-ray flux in the last 5.5 years
is evident. Previous works reported that the electron temperature of the X-ray 
emitting plasma, measured by a single shock model, was decreasing from $kT$ 
$\sim$ 3 keV (day $\sim$4600) to $kT$ $\sim$ 2.2 keV (day $\sim$6200)
(Park et al. 2004;2005a). Our latest data indicate an electron temperature of 
$kT$ $\sim$ 1.6 keV as of 2005 July (day $\sim$6700), which confirms the overall 
spectral softening of SNR 1987A. As the overall spectral shape changes 
with the photon statistics significantly improving, the single-shock model, however, 
became unable to adequately fit the recently obtained data. A two-shock model
is more useful approximation than a single-shock model in order to describe the 
observed spectrum by characteristically representing the decelerated and the fast 
shock components (e.g., Park et al. 2004). We thus fit the X-ray spectrum 
observed in each eleven epoch with a two-component plane-parallel shock model 
(Borkowski et al. 2001). The detailed description of the two-shock spectral 
modeling of SNR 1987A can be found elsewhere (Park et al. 2005c). 

In the spectral fits, we fix some elemental abundances (relative to solar) at the 
values appropriate for the inner ring and the LMC ISM because contributions of the
line emission from those species in the observed energy range are negligible: 
i.e., He (= 2.57), C (= 0.09) (Lundqvist \& Fransson 1996), Ca (= 0.34), Ar (= 0.54), 
and Ni (= 0.62) (Russell \& Dopita 1992). For other species of N (= 0.76), O (= 
0.094), Ne (= 0.29), Mg (= 0.24), Si (=0.28), S (= 0.45), and Fe (= 0.16), we use 
the abundances measured with the spectral analysis of our deep LETG observations 
(Zhekov et al. 2005b), because we believe that the high resolution dispersed spectrum 
obtained from the deep LETG observations provides the best information on the 
elementral abundances. The best-fit models then indicate the electron temperatures 
of $kT$ = 0.22$-$0.31 keV and 2.2$-$3.2 keV for the soft and the hard 
components, respectively. The soft component is most likely in collisional ionization 
equilibrium (ionization timescale $n_et$ $\sim$ 10$^{13}$ cm$^{-3}$ s), while the 
hard component is in non-equilibrium ionization condition ($n_et$ $\sim$ 2 $\times$ 
10$^{11}$ cm$^{-3}$ s). The best-fit total foreground column is $N_H$ = 2.35 $\times$ 
10$^{21}$ cm$^{-2}$. 

\begin{figure*}
\centering
\includegraphics[width=0.9\linewidth]{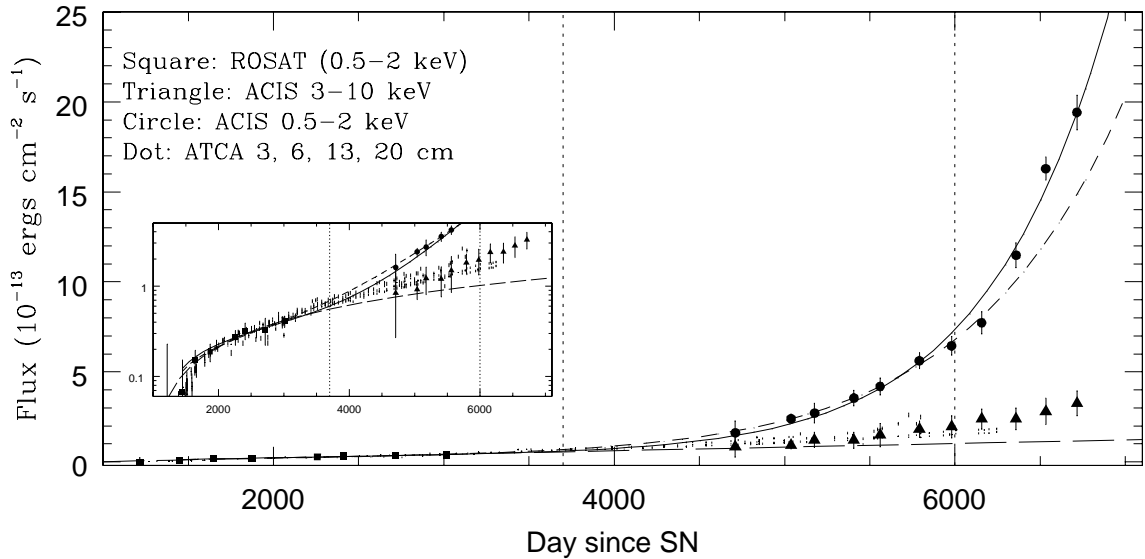}
\vspace{-0.5cm}
\caption{Composite light curves of SNR 1987A (as presented in Park et al. 2005b).
The ROSAT data are taken from Hasinger et al. (1996). The radio fluxes were
obtained with the Australian Telescope Compact Array (ATCA) (provided by L. 
Staveley-Smith) and arbitrarily scaled for the purpose of display. 
The solid curve is the best-fit model to fit entire soft X-ray (0.5$-$2 keV)
light curve (Park et al. 2005b). The short-dashed curve is the best-fit
model from Park et al. (2004), which is extrapolated after day 6000. 
The long-dashed line is a linear fit to the ROSAT data (Burrows et al. 2000),
which is extrapolated to day 7000. The inset is the same light curve plot in 
a log-scale. 
\label{fig:lc}}
\end{figure*}

Based on the two-shock model, we present the X-ray light curves of SNR 1987A
in Figure~3. The soft X-ray light curve shows that the observed flux significantly
deviates from the extrapolation of the simple model used in previous works 
(the short-dashed curve) after day $\sim$6200. We interpret that this up-turn
of the soft X-ray flux is caused by the shock recently beginning to interact with 
the entire inner ring rather than only with small protrusions of the CSM. In fact, 
the modified model (the solid curve), which considers the X-ray flux separately 
before and after the shock interacts with the dense CSM in order for a better 
description of such a condition, can successfully fit the overall soft X-ray light 
curve (Park et al. 2005b).

The light curves of the fractional contributions from the soft and the hard
components to the observed 0.5$-$2 keV flux are presented in Figure~4. The
contribution from the soft component was small and then continues to
increase for the last five years to become dominant since day $\sim$6200.
This long-term change in the fractional light curves suggests that the 
decelerated portion of the shock front by the interaction with the dense CSM
has continuously increased and now the shock front is for the most part decelerated
by the dense inner ring. The soft X-ray light curve and the fractional flux
variations thus consistently support that the blast wave shock began to interact
with the entire inner ring since day $\sim$6200.


%


\section{Radial Expansion Rate}

The radial expansion of the overall X-ray remnant of SN 1987A has been
reported (Park et al. 2002;2004). The expansion rate was estimated to be
$\sim$4200 km s$^{-1}$ until day $\sim$5800 (Park et al. 2004). With the
latest data, we perform X-ray measurements of the radial expansion rate of 
SNR 1987A using a more sophisticated image modeling than that by Park et al. 
(2002;2004). The detailed description of our image analysis for the expansion
measurements can be found elsewhere (Racusin et al. 2005). We find a ``break'' 
in the radial expansion rate at around day 6200 where the rate becomes 
significantly lower from $\sim$3800 km $^{-1}$ to $\sim$1600 km s$^{-1}$
(Figure~5). The deceleration of the radial expansion rate of SNR 1987A is
perhaps expected considering our picture of the blast wave everntually
entering the dense CSM all around the inner ring. It is remarkable that the
predicted deceleration of the expansion rate occurs at day $\sim$6200
which is coincident with the results from the soft X-ray light curves. The
radial expansion rate of SNR 1987A is thus self-consistently supportive of
the shock beginning to interact with the entire inner ring at day $\sim$6200.
 
\begin{figure}
\centering
\includegraphics[angle=-0,width=\linewidth]{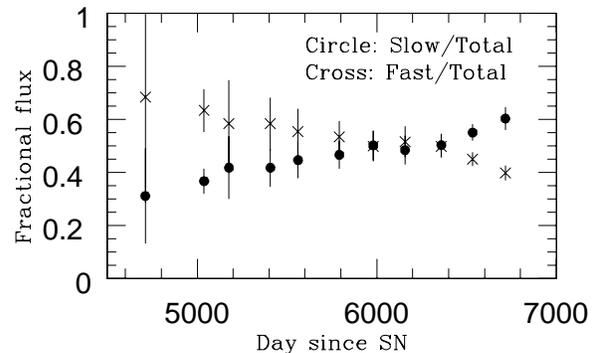}
\vspace{-0.8cm}
\caption{The 0.5$-$2 keV band fractional light curves of SNR 1987A based on the
two-shock model (as presented in Park et al. 2005b).
\label{fig:frac}}
\end{figure}
\begin{figure}
\centering
\includegraphics[angle=-0,width=0.80\linewidth]{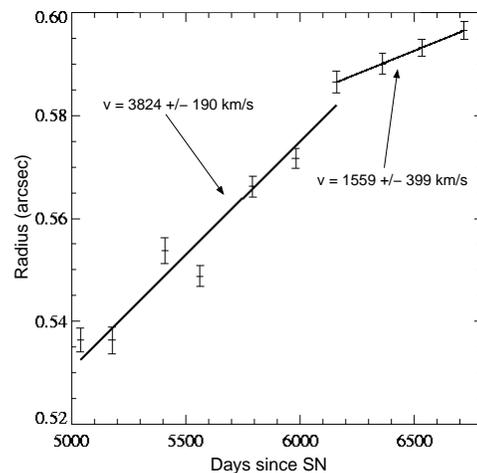}
\caption{The radial expansion rate of SNR 1987A.
\label{fig:rad}}
\vspace{-0.15cm}
\end{figure}
\section{Dispersed Spectrum}
The dispersed X-ray spectrum of SNR 1987A obtained with deep LETG observations 
is presented in Figure~6. The fine structures of X-ray lines are clearly resolved 
with good photon statistics. The detailed discussion of the results from these LETG 
data can be found in Zhekov et al. (2005a;2005b). We note that we chose the roll-angle 
of the observations in order to align the dispersion axis in the north-south direction. 
With this observation setup, assuming that the bulk of X-ray emission originates 
from the interaction between the blast wave and the inner ring, we expected that 
the Doppler shifts due to the the orientation of the inner ring (an inclination 
angle of $\sim$45$^{\circ}$ with the north side toward the Sun) would result in 
systematic distortions of the dispersed images of SNR 1987A: i.e., the overall 
SNR image should be compressed in the negative ($m$ = $-$1) arm while it is 
stretched in the positive ($m$ = +1) arm. The measured line widths indeed reveal 
systematically larger broadenings in the positive arm than those measured in 
the negative arm (Figure~7). These line profiles provide the first direct evidence 
that the X-ray emission is originating from a ``plane'' containing the inner ring
rather than from a spherical volume. These line broadenings imply shock velocities
of v $\sim$ 340$-$1700 km s$^{-1}$.

\begin{figure}
\centering
\includegraphics[angle=-90,width=\linewidth]{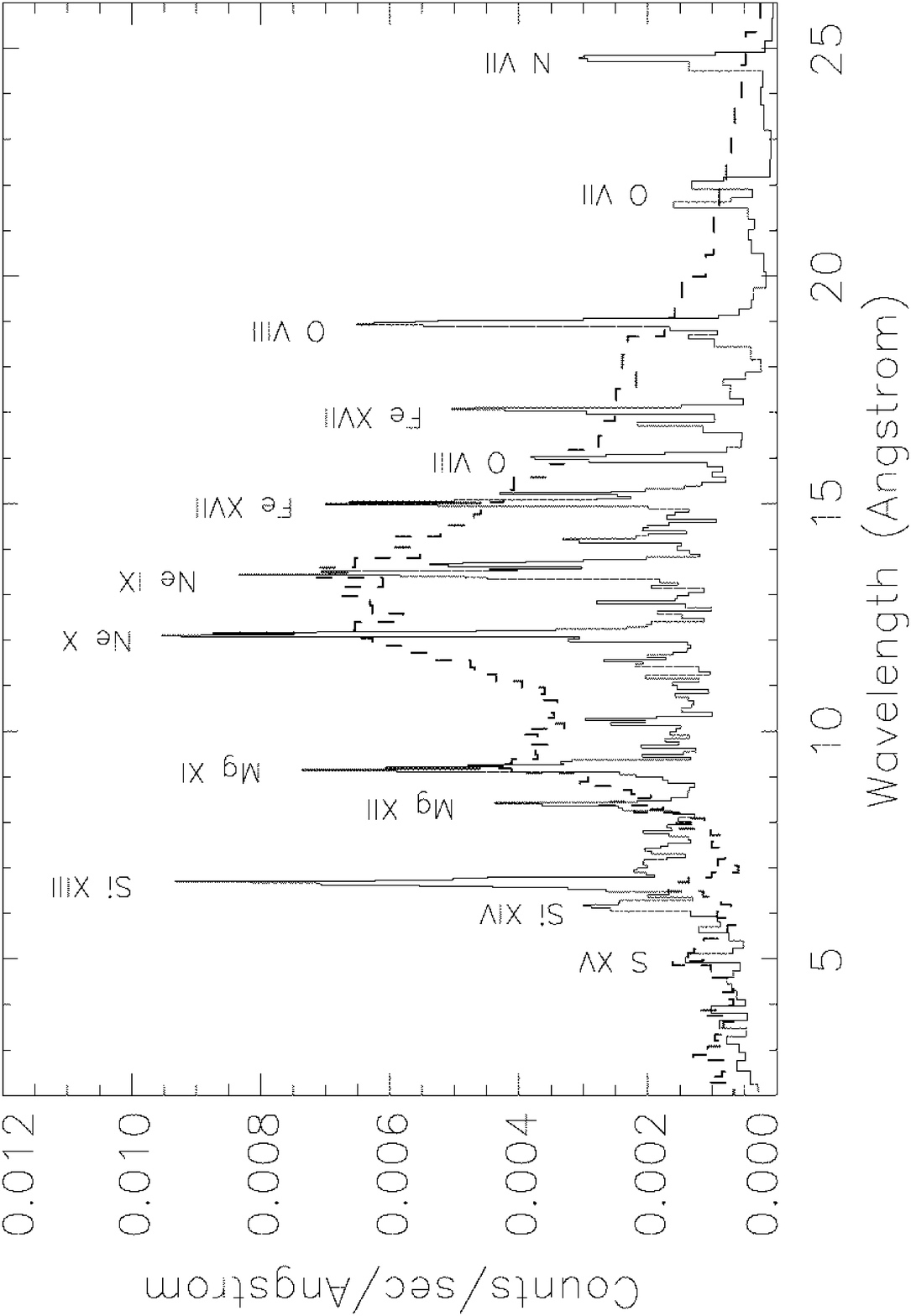}
\caption{The first-order (+1, solid line) and the zeroth-order (dashed line)
spectrum of SNR 1987A from the LEGT observations (as presented in Zhekov et al.
2005a). Some identified lines from key elemental species are marked. 
\label{fig:lineprof}}
\end{figure}

\begin{figure}
\centering
\includegraphics[angle=-90,width=\linewidth]{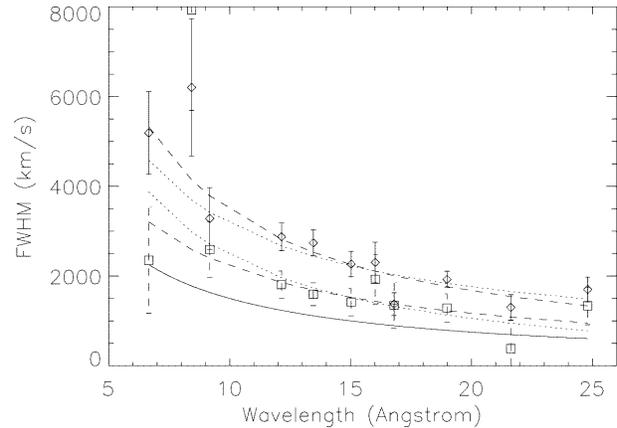}
\caption{Measured line widths (FWHM) for the ``+1'' (diamonds) and the ``$-$1''
(squares) LETG arms (as presented in Zhekov et al. 2005a). The solid curve presents 
the resolving power of the LETG. The dashed and dotted curves are best-fit 
line-broadening parameters for the cases with and without shock stratification, 
respectively (see Zhekov et al. 2005a for the details).
\label{fig:lineprof}}
\end{figure}

The deep LETG observations allow us to directly measure the line flux
ratios from various X-ray lines of SNR 1987A (Zhekov et al. 2005a). In Figure~8,
we present the ranges of the electron temperature ($kT$) and the ionization 
timescale ($n_et$) derived from the plane-parallel shock model, which agree with 
the measured ``G-ratios'' of the He$\alpha$ triplets ($G = [f + i] / r$, where 
f, i, and r are the forbidden, intercombination, and resonance line intensities, 
respectively) and the He$\alpha$/Ly$\alpha$ ratios from O and Si. 
Broad ranges of the electron temperature ($kT$ $\sim$ 0.1$-$2 keV) and the
ionization timescale ($n_et$ $\sim$ 10$^{11}$$-$10$^{13}$ cm$^{-3}$ s) are
consistent with the measured line ratios.

\begin{figure*}
\centering
\includegraphics[angle=-90,width=0.45\linewidth]{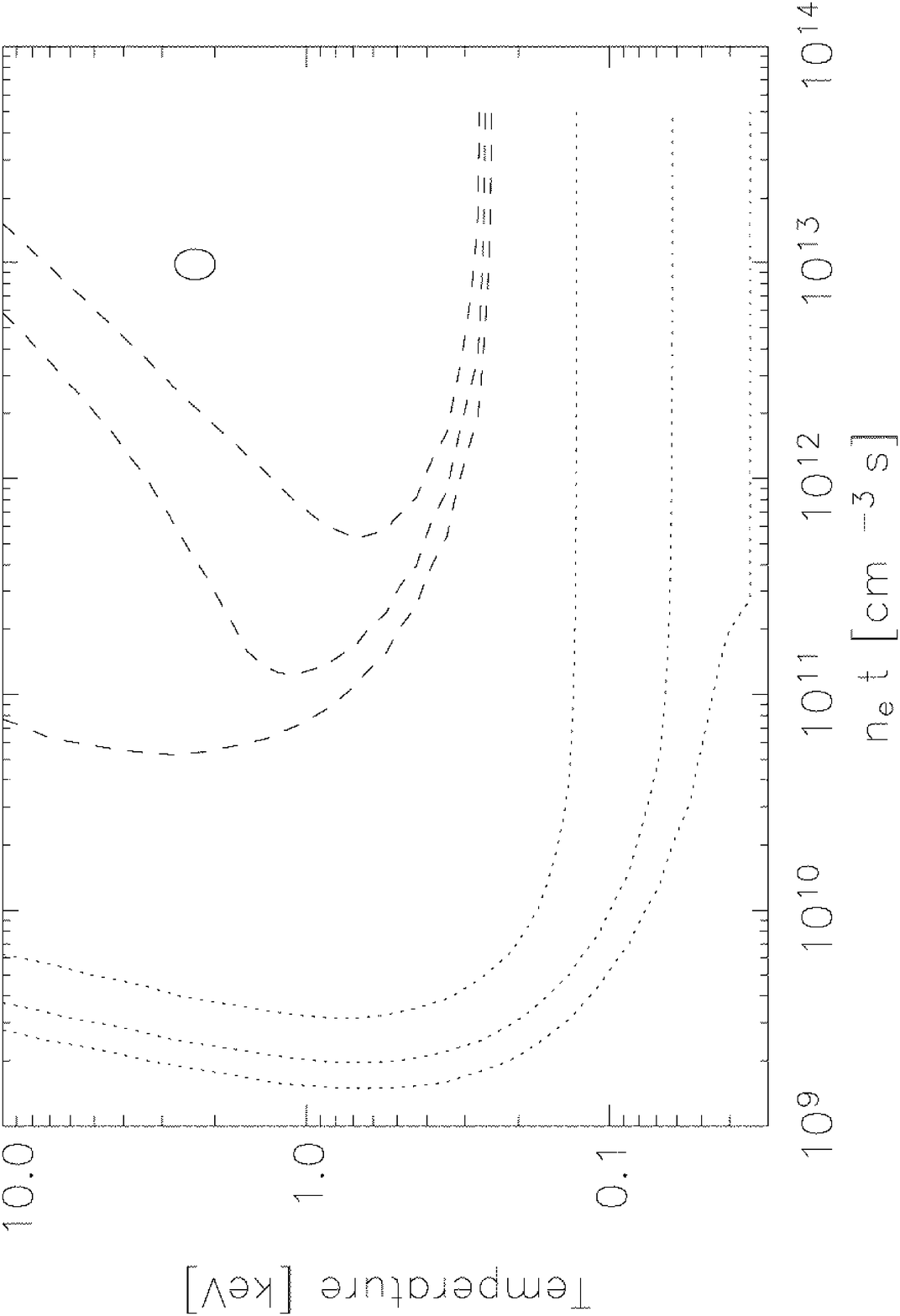}
\includegraphics[angle=-90,width=0.45\linewidth]{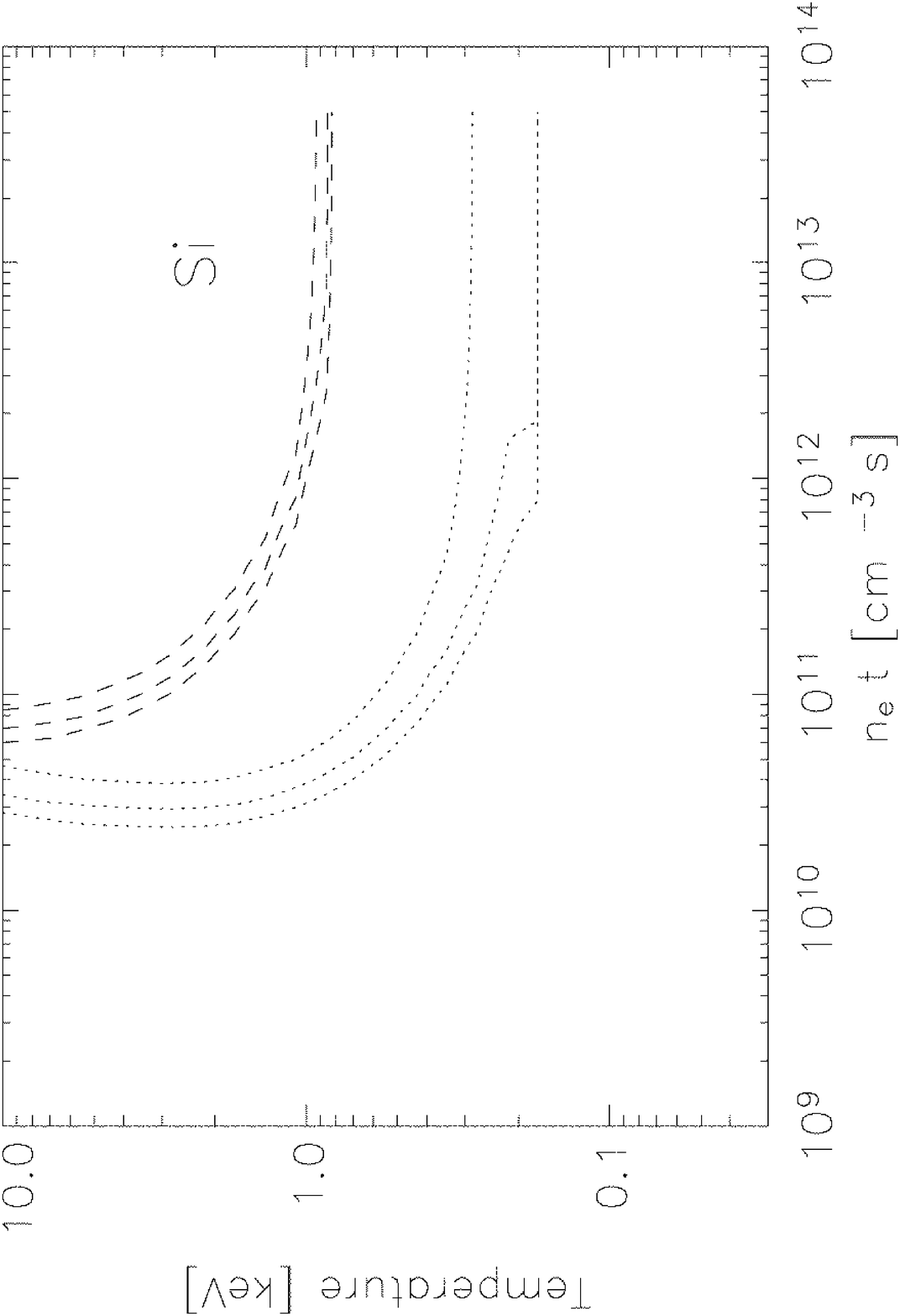}
\caption{Allowed $kT$ vs. $n_et$ ranges to match the measured G-ratios (dotted
curves) and He$\alpha$/Ly$\alpha$ ratios (dashed curves) of O (left panel) and 
Si (right panel) lines from SNR 1987A (as presented in Zhekov et al. 2005a). 
The three curves for each ratio correspond to the measurement and $\pm$1$\sigma$ 
uncertainties, respectively.  
\label{fig:lineprof}}
\end{figure*}

\section{Discussion}

The bulk of soft X-ray emission of SNR 1987A most likely originates from the 
interaction between the blast wave shock and the clumpy structure of the 
pre-existing (before the SN) dense equatorial CSM. The emergence of the first 
optical spot in day $\sim$3700 indicated the beginning of such interactions. 
Days $\sim$6000$-$6200 appear to mark another important milestone in the evolution 
of SNR 1987A: i.e., the blast wave reaches the dense CSM around the entire inner 
ring. This interpretation is supported by several observations. 

(1) A simple model assuming a single shock propagating into ambient
medium with an exponential density profile has successfully described the soft
X-ray light curve for the previous $\sim$13 years. However, an up-turn of the soft 
X-ray flux since day $\sim$6200 cannot be described with the simple
constant-velocity shock model. A significant contribution from the decelerated shock 
needs to be considered in order to adequately fit the data (Park et al. 2005b).
(2) The fractional contribution to the observed 0.5$-$2 keV flux from the soft 
component of the two-shock model has continuously increased. The soft component
represents the X-ray emission from the decelerated shock and becomes dominant 
since days $\sim$6000$-$6200 (Park et al. 2005b).
(3) The X-ray radial expansion rate significantly reduces since day $\sim$6200
(Racusin et al. 2005).
(4) The shock kinematics obtained by the LETG observations (at day $\sim$6400) 
indicate velocities of $\sim$340$-$1700 km s$^{-1}$ for the X-ray emitting plasma
(Zhekov et al. 2005a). These velocities are significantly lower than those estimated 
from earlier X-ray and radio images (v $\sim$ 3000$-$4000 km s$^{-1}$).
(5) The soft X-ray images indicate that the brightening has become {\it global}
since day $\sim$6000 (Park et al. 2005b).
(6) The optically bright spots become prevailing over the entire inner ring by 
day $\sim$6000 (e.g., Plate 1 in McCray 2005).

On the other hand, the hard X-ray light curve shows significantly lower flux
increase rate (Figure~3). It is remarkable that the hard X-ray light curve is 
similar to that of the radio data. The radio emission appears to primarily 
originate from the synchrotron emission behind the reverse shock (Manchester et al. 
2005). The similarity between the radio and hard X-ray light curves might suggest 
that the hard X-rays originate from the same population of electrons to produce 
radio synchrotron emission. However, a better correlation of the radio image with 
the hard X-ray image than with the soft X-ray image is not clear (Park et al. 2005b). 
Thus, the lower flux increase rate in the hard X-ray band might simply be caused by 
the continuous softening of the X-ray spectrum due to the shock interaction with 
the dense CSM. Follow-up monitoring of the light curve, the morphology, and the
spectral properties in the hard X-ray band will be essential to unveil the true 
origin of the hard X-ray emission of SNR 1987A.

With the recent deep LETG observations, we for the first time measure the
individual X-ray line profiles from SNR 1987A (Zhekov et al. 2005a). The 
systematic difference in the line broadening between the positive and negative 
dispersion arms provide direct observational evidence for the X-ray emission 
originating in the equatorial plane of the inner ring. More surprisingly, 
the derived velocity (v $\sim$ 340$-$1700 km s$^{-1}$) of the X-ray emitting 
plasma is significantly lower than the average shock velocities estimated with 
the previous X-ray and radio data (v $\sim$ 3000$-$4000 km s$^{-1}$). The 
low velocity suggests that the blast wave has been significantly decelerated 
by the dense inner ring, which is consistent with the results from the X-ray 
light curve and image analyses. 

The X-ray line ratio measurements indicate that the X-ray emitting plasma of 
SNR 1987A cannot be adequately described by a single combination of the electron 
temperature and the ionization timescale. Instead, the observed X-ray lines 
originate from wide ranges of the plasma temperature and the ionization 
age. The multi-phases of the X-ray emitting thermal plasma of SNR 1987A was 
suggested by the undispersed spectrum and is clearly confirmed by the dispersed
spectrum. This plasma structure is most likely due to the interaction of 
the blast wave with the complex density gradients near the boundary of the dense 
inner ring.

\section*{Acknowledgments}

The authors thank L. Staveley-Smith for providing unpublished radio fluxes.
This work was supported in part by SAO under Chandra grants GO4-5072A, GO4-5072B,
and GO5-6073X.



\begin{thebibliography}{}





\bibitem[Borkowski et al.(2001)]{bor01} Borkowski, K. J., Lyerly, W. J., \& Reynolds,
S. P. 2001, ApJ, 548, 820
\bibitem[Burrows et al.(2000)]{bur00} Burrows, D. N. et al. 2000, ApJ, 543, L49
\bibitem[Hasinger et al.(1996)]{has96} Hasinger, G., Aschenbach, B., \& Tru\"mper, J.
1996, A\&A, 312, L9
\bibitem[Manchester et al.(2005)]{man05} Manchester, R. N. et al. 2005, ApJ, 628, L131
\bibitem[McCray(2005)]{mccray05} McCray, R. 2005, \emph{Cosmic Explosions}, Proc. IAU
Colloquium No. 192, ed., J. M. Marcaide and K. W. Weiler, (Heidelberg: Springer), 77
\bibitem[Michael et al.(2002)]{michael02} Michael, E. et al. 2002, ApJ, 574, 166
\bibitem[Park et al.(2002)]{park02} Park, S. et al. 2002, ApJ, 567, 314
\bibitem[Park et al.(2004)]{park04} Park, S. et al. 2004, ApJ, 610, 275
\bibitem[Park et al.(2005a)]{park05a} Park, S. et al. 2005a, AdSpR, 35, 991
\bibitem[Park et al.(2005b)]{park05b} Park, S. et al. 2005b, ApJL, in press (astro-ph/0510442)
\bibitem[Park et al.(2005c)]{park05c} Park, S. et al. 2005c, in preparation
\bibitem[Racusin et al.(2005)]{rac05} Racusin, J. L. et al. 2005, in preparation
\bibitem[Zhekov et al.(2005a)]{zhekov05a} Zhekov, S. A. et al. 2005a, ApJ, 628, L127
\bibitem[Zhekov et al.(2005b)]{zhekov05b} Zhekov, S. A. et al. 2005b, in preparation

\end{thebibliography}
\end{document}